\documentclass[sigconf]{acmart}

\usepackage{algorithmic}
\usepackage{graphicx}
\usepackage{textcomp}
\usepackage[most]{tcolorbox}
\usepackage{booktabs}
\usepackage{tabularx}
\usepackage{makecell}

\usepackage{tikz}
\usetikzlibrary{arrows.meta}
\usetikzlibrary{positioning}

\AtBeginDocument{%
  }

\setcopyright{acmlicensed}
\copyrightyear{2026}
\acmYear{2026}
\acmDOI{XXXXXXX.XXXXXXX}
\acmConference[Gamify '26]{the international workshop on Gamification in Software Development, Verification, and Validation}{June 09--12, 2026}{Glasgow, UK}
\acmISBN{978-1-4503-XXXX-X/2018/06}
\begin{document}

\title{Gamifying Architectural Governance to Reduce Organizational Coupling in Microservice Systems}

\author{Xiaozhou Li}
\affiliation{%
  \institution{Free University of Bozen-Bolzano}
  \city{Bolzano}
  \country{Italy}}
\email{xiaozhou.li@unibz.it}

\begin{abstract}

Microservice is a popular software architecture that relies on decentralized teams and clear service ownership to support modularity and scalability. 
However, in practice, developers frequently contribute across multiple services, creating organizational coupling (OC) that gradually erodes architectural boundaries and increases coordination overhead. This study proposes a vision for behavior-driven architectural governance through gamification in microservice systems to influence developer behavior and reduce OC. 
Our approach introduces a gamified framework that continuously mines repository data to detect architectural boundary violations and increasing service dependencies, and translates those signals into gameful designs, including points, badges, leaderboards, and architecture improvement quests. We outline a conceptual framework that integrates repository mining, architectural metrics, and gamification mechanisms to encourage developers to maintain service boundaries and improve architectural maintainability. Furthermore, we present an evaluation roadmap to assess the impact of gamified OC governance and developer engagement. This work aims to open a new research direction at the intersection of software architecture governance, socio-technical analysis, and gamification, highlighting the potential of behavioral incentives to support sustainable microservice evolution.

\end{abstract}




\maketitle


\section{Introduction}
\label{sec:introduction}


Microservice architectures enable modular, scalable, and independently deployable systems \cite{dragoni2017microservices}. 
However, this decomposition introduces a key socio-technical challenge: aligning system modularity with team organization. According to Conway’s Law, system structures inevitably reflect the communication patterns of the organizations that build them \cite{conway1968committees}. In microservice ecosystems, this implies that the service boundaries should ideally align with team responsibilities. However, in practice, such an alignment is difficult to sustain. Empirical studies show that developers frequently contribute across multiple services, violating the principle of "one microservice per team/developer" and introducing OC, i.e., hidden dependencies created by cross-service contributions \cite{amoroso2023one,li2023evaluating}. 
Such coupling increases the coordination cost, blurs ownership, and erodes modularity. At the same time, teams exhibit varying levels of cohesion, reflecting how focused and stable their contributions are within a service \cite{li2025exploring}.


Existing approaches for architectural governance in microservice systems primarily focus on monitoring, analyzing, and tooling \cite{wang2023operation,schneider2025comparison}. Although these techniques provide valuable information about architectural evolution, they remain largely passive diagnostic mechanisms. Although recent research has also proposed quantitative metrics to capture organizational coupling and cohesion, they remain largely diagnostic: they measure socio-technical misalignment, but do not actively influence developer behavior \cite{li2023evaluating}. This creates a critical gap between observability and governance. Providing metrics is often insufficient to drive behavioral change in complex, distributed development environments. Notably, not all cross-service contributions indicate undesirable coupling when some coordination can be necessary and beneficial, e.g., coordinated refactoring and platform-wide upgrades. The governance goal is therefore not to eliminate cross-service activity, but to distinguish persistent, unjustified boundary erosion from legitimate collaborative work, and to intervene only in the former case.

In this paper, we argue that gamification provides a promising mechanism to bridge this gap. By embedding architectural feedback into interactive and motivational systems, gamification can transform passive measurements into active behavioral interventions. Rather than treating architectural governance as a top-down control process, gamification enables continuous, bottom-up alignment by nudging developers toward desirable contribution patterns. We propose a \textit{Gamified Architecture Governance Framework} that integrates organizational coupling and cohesion metrics with gamification mechanisms such as points, badges, leaderboards, and behavioral nudges. The framework establishes a continuous feedback loop between developer activity and architectural quality, encouraging behaviors that promote high cohesion and low coupling at the organizational level.


This paper makes the following contributions: 1) conceptualize organizational coupling and cohesion as governable behavioral phenomena in microservice development; 2) propose a gamified governance framework that links socio-technical metrics to behavioral incentives; and 3) outline a research agenda for evaluating and operationalizing gamified interventions in software architecture governance. Generally, by shifting from passive measurement to active behavioral shaping, this work opens a new direction for managing socio-technical complexity in modern software systems.


The remainder of this paper is structured as follows. Section \ref{sec:background} provides background and motivation. 
Section \ref{sec:framework} presents the gamified architectural governance framework. Section \ref{sec:evaluation} outlines an evaluation plan 
Section \ref{sec:discussion} discusses potential research directions and threats to validity, and Section \ref{sec:conclusion} concludes the paper.
\section{Background and Motivation}
\label{sec:background}


Microservice architectures decompose systems into small, independently deployable services aligned with specific business capabilities, enabling scalability and team autonomy when service boundaries and ownership are well defined \cite{dragoni2017microservices}. 
However, in practice, this alignment often degrades as developers contribute across multiple services, creating OC, a phenomenon where shared or repeated cross-service contributions blur ownership boundaries, increase coordination overhead, and erode modularity \cite{li2023evaluating,mani2026organizational}. 
Previous work shows that such patterns can be detected through repository mining by analyzing contribution histories and cross-service commits, enabling continuous monitoring of OC in microservice systems \cite{li2023evaluating,li2025exploring,mani2026organizational}.


To address architectural degradation, several approaches have been proposed, including dependency analysis, logical coupling detection, and architectural metrics derived from commit history \cite{li2023analyzing,bakhtin2024temporal,abdelfattah2023microservice,amoroso2024understanding}. Although these techniques provide valuable insights into structural and evolutionary properties, they remain primarily diagnostic: they identify issues but do not directly influence the behaviors that cause them. As a result, architectural governance remains largely reactive, often detecting problems only after degradation has occurred.

Meanwhile, gamification, the use of game design elements such as points, badges, leaderboards, and challenges in non-game contexts \cite{deterding2011game}, has been widely studied as a mechanism to motivate engagement and behavioral change \cite{hamari2014does}. In software engineering, gamification has been applied to tasks such as bug reporting, code review participation, developer onboarding, and learning platforms \cite{khandelwal2017impact,arai2014gamified,ruiz2024gamification}. These approaches demonstrate that both intrinsic and extrinsic incentives can effectively influence developer behavior and increase participation in otherwise overlooked activities.


Despite these advances, the application of gamification to software architecture governance remains largely unexplored. Architectural quality in microservice systems is shaped by everyday development decisions, including how developers distribute contributions across services and manage dependencies. Encouraging developers to maintain architectural boundaries is therefore fundamentally a behavioral challenge, not solely a technical one \cite{ali2018architecture}.




Architectural governance differs from tasks like bug fixing or code review in that it is indirect and lacks immediate feedback. Gamification is well suited to this context because it makes otherwise invisible architectural signals visible and actionable, providing timely, behavior-oriented feedback that can align daily development activities with long-term architectural goals. Compared with existing approaches, this vision 1) shifts from purely diagnostic analysis to behavior-oriented intervention, and 2) complements policy-driven mechanisms through feedback, incentives, and developer engagement. Unlike policy-based governance, e.g., CODEOWNERS \cite{lulla2025automated}, which enforces ownership and compliance rules, gamification acts as a softer behavioral intervention. Furthermore, in contrast to prior gamification work focused on task-level activities (e.g., bug handling, code review, onboarding), this framework targets architectural behavior as the primary unit of intervention. 

\section{Gamified Architectural Governance Framework}
\label{sec:framework}

\begin{figure*}[!ht]
    \centering
    \includegraphics[width=\textwidth]{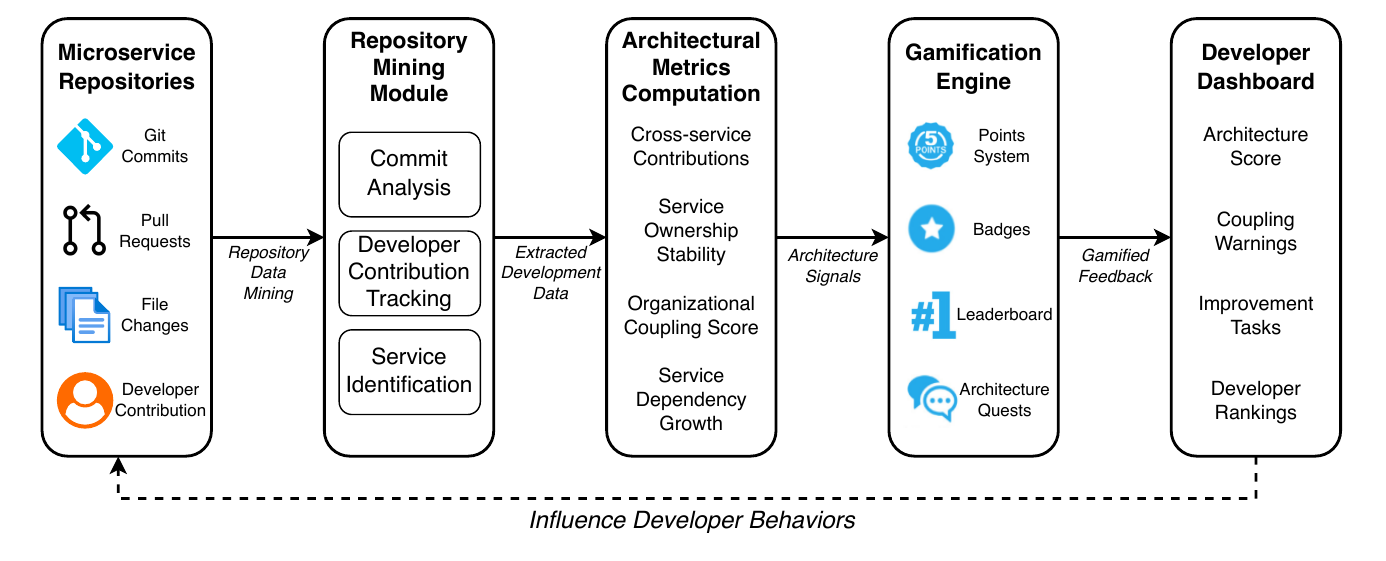} 
    \caption{A closed-loop gamified governance framework that transforms repository-derived architectural signals into behavioral feedback to influence developer actions and improve socio-technical alignment.}
    \label{fig:framework}
\end{figure*}

To bridge measurement and intervention, we propose a Gamified Architecture Governance Framework integrating socio-technical metrics with behavioral feedback.
The framework forms a closed loop that monitors activity, evaluates alignment, and influences behavior.
As shown in Figure \ref{fig:framework}, the framework consists of four main components: (1) data collection, (2) socio-technical analysis, (3) gamification engine, and (4) feedback and adaptation. Figure 1 illustrates the overall architecture of the framework.
At a high level, the framework transforms raw development activity into actionable feedback that encourages developers to align their behavior with architectural goals.

\subsection{Data and Analysis Layer}

The data layer collects development activity from repositories (e.g., GitHub), including commits, file changes, and contributor interactions. 
From these data, the framework computes OC, capturing cross-service contribution dependencies, and Organizational Cohesion, capturing the focus and stability of team contributions within services.
These metrics provide a quantitative representation of socio-technical alignment. High coupling indicates potential coordination risks, while high cohesion reflects stable and well-defined ownership.

Cross-service work is interpreted contextually rather than uniformly penalized. Necessary contributions arise from coordinated refactoring, shared infrastructure changes, incident response, or intentionally coupled releases, whereas harmful coupling is reflected in repeated ad hoc switching, chronic ownership overlap, or persistent cross-service activity without architectural intent. This distinction enables governance to target boundary erosion rather than legitimate collaboration.



To ensure fair interpretation across projects, the framework uses project-relative thresholds and normalized signals instead of fixed cutoffs. OC, contribution switching, and cohesion are computed over rolling time windows and normalized by project size, service count, contributor count, and recent activity. Interventions are thus triggered by deviations from expected baseline behavior rather than raw activity alone.


\subsection{Gamification Engine}





The gamification engine transforms architectural signals into behavioral interventions. Instead of passively presenting metrics, it operationalizes organizational coupling and cohesion as actionable feedback to guide developer decisions in everyday work.

It follows a closed-loop model in which 1) developer actions generate architectural signals, 2) these signals are interpreted as performance indicators, and 3) targeted feedback influences subsequent actions. Grounded in goal-setting and self-determination theories, the design balances extrinsic incentives with intrinsic motivation, emphasizing feedback, mastery, and team improvement over purely competitive rewards, thus reducing superficial compliance and motivation crowding-out.

\subsubsection{Multi-Level Gamification Design}

To address the socio-technical nature of microservice development, the gamification engine operates at both individual and team levels, combining short-term feedback with long-term engagement. We define four complementary intervention types:

\textbf{Points and Scoring}. Developers receive scores based on contribution patterns relative to architectural goals. Actions that improve cohesion (e.g., sustained service focus) or reduce coupling (e.g., limited cross-service switching) are rewarded, while excessive boundary violations lower scores. Scoring remains context-aware to avoid penalizing necessary cross-service work.

\textbf{Badges and Progression}. Badges recognize sustained architecture-aligned behavior (e.g., \emph{Service Specialist}, \emph{Boundary Guardian}), fostering progression and mastery, and reinforcing long-term consistency over short-term optimization.

\textbf{Leaderboards and Social Transparency}. Leaderboards expose architectural alignment at individual and team levels, promoting awareness and lightweight competition. To mitigate negative effects, rankings can be scoped (e.g., team-level) and complemented with collaborative metrics, emphasizing collective improvement. In practice, team-scoped or opt-in leaderboards are preferred over global rankings.

\textbf{Behavioral Nudges and Real-Time Feedback}. The system issues contextual nudges when harmful patterns emerge. For instance, repeated short-term cross-service contributions may trigger suggestions to refocus or coordinate with responsible teams. These just-in-time interventions make otherwise invisible architectural risks actionable.

\subsubsection{Architecture Improvement Quests}

Beyond continuous feedback, we introduce architecture improvement quests as a key extension to traditional gamification. Quests define explicit, goal-oriented challenges aligned with architectural objectives, such as reducing coupling between specific services, stabilizing ownership of a service, or refactoring cross-service dependencies.

Quests can be assigned at individual or team levels and are time-bound, encouraging coordinated efforts toward measurable improvements. This mechanism shifts gamification from passive feedback to active goal-driven intervention, allowing teams to collectively address architectural issues.

\subsubsection{Design Considerations and Safeguards}

To ensure effectiveness and avoid unintended consequences, the gamification engine incorporates several safeguards: \textbf{Context sensitivity}: distinguishing necessary from harmful cross-service contributions through project and change context; \textbf{Balanced incentives}: combining rewards and nudges without over-reliance on competition, and prioritizing sustained improvement over short-term score maximization; \textbf{Transparency}: clearly communicating how metrics and rewards are computed, and allowing developers to understand why feedback is triggered; and \textbf{Adaptability}: allowing customization based on project context and team structure.


By embedding these mechanisms into development workflows, the gamification engine transforms architectural governance from external monitoring into an integrated behavior-aware system. It enables continuous alignment between developer actions and architectural intent, supporting the sustainable evolution of microservice systems. Unlike prior gamification approaches in software engineering that focus on task-level engagement, this approach targets architectural behavior as the unit of intervention, enabling direct influence on socio-technical system structure.

\subsection{Feedback Loop and Behavioral Adaptation}

A key feature of the framework is its continuous feedback loop. Developer actions influence metrics, which are then translated into gamified feedback that, in turn, shapes future behavior. 
For example, repeated cross-service contributions increase coupling metrics, triggering nudges or reduced rewards that encourage developers to refocus their contributions.
Over time, this loop promotes self-regulation, where developers internalize architectural goals and adjust their behavior accordingly.

\subsection{Expected Outcomes}

The framework aims to achieve: 1) Reduced organizational coupling through minimized cross-service dependencies, 2) Increased team cohesion via focused and stable contributions, and 3) Improved alignment between architectural design and organizational structure.
By embedding governance into everyday development workflows, the framework shifts architecture management from reactive monitoring to proactive behavioral guidance.
\section{Evaluation Plan}
\label{sec:evaluation}

As this paper presents a vision for gamified architectural governance, we outline an evaluation strategy aimed at assessing both the feasibility of the approach and its potential impact on OC and developer behavior. The evaluation is structured combining repository mining, simulation, and human-centered studies.

\subsection{Data Source}

The framework relies on repository data from microservice-based systems. We use publicly available open-source projects, which provide rich histories for analyzing developer behavior and architectural evolution. Candidate systems include \textit{SockShop}, \textit{sShop}, and \textit{Spinnaker}. Although benchmark systems support early feasibility analysis, later validation should focus on larger, long-lived projects with richer socio-technical histories to improve ecological validity. 
From these repositories, we collect commit data (author, timestamp, modified files) to reconstruct contributions, file paths, and structure to identify service boundaries, pull request metadata for collaboration, and dependency information to analyze service interactions. These data enable computation of architectural metrics, including cross-service contributions and OC, as shown in previous work.

\subsection{Baseline Analysis of Architectural Behavior}

In the first phase, we establish a baseline understanding of architectural behavior in microservice systems through repository mining. We compute the key metrics, including: Cross-service contribution ratio, Service ownership stability, OC scores, and Contribution switching patterns. This analysis provides insights into boundary-crossing frequency and OC evolution. It also enables the identification of typical coupling patterns and hot-spots. These observations also provide the basis for comparison against two non-gamified baselines: a transparent-metrics-only condition and policy-driven governance mechanisms, e.g., ownership enforcement rules. The comparison between transparent metrics only and gamified feedback isolates the effect of gamification beyond visibility, while comparison with policy-driven governance evaluates whether behavioral incentives provide additional benefits over rule-based enforcement.

\subsection{Simulation of Gamified Governance}

In the second phase, we evaluate gamification impact through simulation. Using mined repository data, we model how developer behavior may change under different gamification strategies.
Specifically, we simulate scenarios based on a simple behavior model calibrated with pilot observations or prior assumptions: 1) rewards for reducing cross-service contributions, 2) penalties for excessive boundary violations, and 3) quests incentivizing refactoring and improved ownership. Outcomes are compared to baseline conditions, assessing reductions in OC, improvements in ownership stability, shifts in contribution distribution, and architectural health proxies such as dependency complexity and coupling hot-spots.
This phase enables exploration of gamification effectiveness before deployment in real environments.

\subsection{Developer-Centered Evaluation}

The final phase focuses on human-centered evaluation through controlled studies with: 1) graduate students in microservice projects, 2) open-source contributors, and 3) industry practitioners (when accessible).

Participants will use a prototype of the gamified governance framework. The evaluation measures: 1) behavioral changes (e.g., reduced cross-service contributions), 2) perceived usefulness and usability, 3) motivation and engagement, and 4) architectural health proxies, such as dependency complexity, logical coupling trends, and coupling hot-spots. Where feasible, studies will compare gamified and non-gamified conditions using randomized or quasi-experimental designs (e.g., A/B testing, pre/post), including team-level quests and opt-in, team-scoped leaderboards.

Qualitative feedback will capture impacts on decision-making, collaboration, and perceived fairness, as well as potential unintended effects such as excessive competition or stress.
\section{Discussion}
\label{sec:discussion}

\subsection{Research Agenda}

This vision opens research directions at the intersection of software architecture, socio-technical systems, and gamification. We outline key research questions:

\textit{RQ1: Which gamification mechanisms are most effective for influencing architecture-aware behavior?} Different elements (e.g., points, badges, leaderboards, quests) may affect motivation differently. Identifying those that effectively promote boundary respect and reduce cross-service contributions is essential. 

\textit{RQ2: Can gamification reduce OC in microservice systems?} 
A central question is whether gamified feedback can measurably decrease OC metrics, such as cross-service contributions and contribution switching. Longitudinal studies are needed to assess whether such improvements are sustained over time.

\textit{RQ3: How does gamified governance affect architectural quality and maintainability?}
Beyond behavioral changes, it is important to evaluate whether reduced OC translates into improved architectural outcomes, such as lower dependency complexity, fewer architectural violations, and improved maintainability.

\textit{RQ4: What is the impact of gamification on developer experience and collaboration?} 
Gamification can influence not only individual behavior but also team dynamics. Understanding its effects on motivation, collaboration, and developer satisfaction is critical to ensure that incentives do not introduce negative side effects.

\textit{RQ5: How can gamification be adapted to different project contexts?} 
Microservice systems vary in scale, domain, and organizational structure. Future work should explore how gamified governance mechanisms can be tailored to different contexts, including open-source and industrial environments.

\subsection{Threat to Validity}

The proposed metrics, such as cross-service contributions and OC, may not capture all forms of architectural violations. Some cross-service contributions may be necessary or beneficial, and distinguishing harmful from acceptable coupling remains a challenge. 

In simulation-based evaluation, assumptions about developer behavior may not fully reflect real-world decision-making. Developers may respond differently to gamified incentives than predicted, potentially limiting the accuracy of simulated outcomes.

The use of open-source microservice projects can limit the generalizability of the results to industrial settings, where team structures, development processes, and constraints differ significantly.

Gamification can introduce unintended consequences, such as excessive competition, gaming of the system, or increased stress among developers. Poorly designed incentives may encourage superficial compliance rather than meaningful architectural improvements.

Integrating gamified governance into existing development workflows may face resistance from practitioners, particularly if perceived as intrusive or misaligned with team goals. Ensuring usability and alignment with developer practices is therefore critical.
\section{Conclusion}
\label{sec:conclusion}

This paper presents a vision for gamifying architectural governance in microservice systems. We proposed a conceptual framework that integrates repository mining, architectural metrics, and gamification mechanisms to transform architectural signals into actionable, behavior-oriented feedback. By introducing game-based incentives such as points, badges, and quests, the framework aims to actively encourage developers to respect service boundaries and reduce organizational coupling. This work highlights a shift from purely analytical governance toward a behavior-driven approach, where architectural quality is supported not only through measurement but also through motivation. By bridging software architecture, socio-technical analysis, and gamification, we aim to open a new research direction for supporting sustainable microservice evolution. Future work will focus on empirically evaluating the effectiveness of gamified governance in real-world settings and refining its design to balance architectural goals with developer experience. This vision advocates a paradigm shift from monitoring the architecture to actively shaping it through developer behavior.


\bibliographystyle{ACM-Reference-Format}
\bibliography{bib}

\end{document}